\pdfoutput=1 %arXiv admin added this line
\documentclass[onecolumn,preprint]{revtex4-1}

\usepackage{graphicx}
\usepackage{dcolumn}
\usepackage{bm}

\begin{document}

\title{Semiclassical and Quantum Analysis of a Free Particle Hermite Wavefunction}

\author{P Strange}
\affiliation{School of Physical Sciences, University of Kent, Canterbury, Kent, CT2 7NH, UK.}

\date{\today}

\begin{abstract}
In this paper we discuss a solution of the free particle Schr\"odinger equation in which the time and space dependence are not separable. The wavefunction is written as a product of exponential terms, Hermite polynomials and a phase. The peaks in the wavefunction decelerate and then accelerate around $t=0$.  We analyse this behaviour within both a quantum and a semi-classical regime. We show that the acceleration does not represent true acceleration of the particle but can be related to the envelope function of the allowed classical paths. Comparison with other "accelerating" wavefunctions is also made. The analysis provides considerable insight into the meaning of the quantum wavefunction.
\end{abstract}

\maketitle

\section{\label{Intro}Introduction:}
The Schr\"odinger equation is at the centre of non-relativistic quantum theory. The usual method of solution is to separate the time and space dependence of the wavefunction and  solve the time independent Schr\"odinger equation\cite{sara}. This can be done analytically for a few simple model potentials and otherwise it is straightforwardly amenable to numerical solution. This approach has had huge success in describing a vast array of physical phenomena, in particular the electronic and structural properties of atoms, molecules and solids.  The Schr\"{o}dinger equation has also been shown to have more exotic solutions such as accelerating Airy wavefunctions\cite{bb, lek}. The wavefunction introduced by Berry and Balazs also has the remarkable property that it does not broaden with time. However it does not represent a single particle because it is not square-integrable and thus is not an element of a Hilbert Space, therefore there is no contradiction with Ehrenfest's theorem. Later Lekner\cite{lek} derived a more general form of the Airy wavefunction and which is both well-behaved and square-integrable. Both the expectation values of position and momentum show no acceleration, except in the Berry-Balazs limit where the wavefunction is not square-integrable. In a subsequent paper Nguyen and Lekner\cite{lek2} were able to derive wavefunctions that describe true acceleration via an extended Galilean transformation of a free particle wavefunction, where the extended Galilean transformation also introduces a potential into the Hamiltonian which drives the acceleration. The Schr\"odinger equation has an identical mathematical form to the paraxial wave equation and so such solutions have been employed as the basis of solutions of Maxwell's equations to describe electromagnetic radiation that can change direction as it propagates\cite{sivi,hac,kam}. In this paper we derive, describe and try to gain insight into an exotic, apparently accelerating solution of the free particle Schr\"odinger equation that is square-integrable and which also displays some unusual characteristics. In section II we write down and describe the solution. In section IIIA we perform a quantum mechanical analysis of the wavefunction and show that the probability density accelerates. Accelerating probability densities for free particles is an apparent contradiction. However they have been found to be explicable on the basis of classical mechanics\cite{bb}. Therefore in section IIIB  we have performed a parallel classical analysis to show how the accelerating probability density can be understood and how the semiclassical and quantum descriptions are related. In section IV we compare the properties of our Hermite wavefunction with those of a Gaussian and the the Airy wavefunction of Lekner\cite{lek}. In conclusion we discuss the considerable insight this calculation yields into the meaning of the quantum mechanical wavefunction. 
\vskip 5mm
\section{Theory}
The time-dependent free particle Schr\"odinger equation in 1+1 dimensions is 
\begin{equation}
i\hbar \frac{\partial \psi (x,t)}{\partial t} =-\frac{\hbar^2}{2m}\frac{\partial^2\psi(x,t)}{\partial x^2}
\label{se}
\end{equation}
Using a symmetry analysis and separation of the variables\cite{mill, kal} it is possible to find a solution of equation (\ref{se}) as
\begin{eqnarray}
\psi (x,t)= & \sqrt{\frac{1}{n!}}  \left(\frac{m}{\hbar t_c\pi}\right)^{1/4} \frac{2^{-n/2}}{(1+t^2/t_c^2)^{1/4}} \times\cr
& \cr
&  e^{imx^2t/(2\hbar(t_c^2+t^2)} e^{-i(n+1/2)\arctan (t/t_c)}\times \cr 
& \cr
& e^{-mt_cx^2/(2\hbar(t_c^2+t^2))} H_n\left(\left(\frac{mt_c}{\hbar(t_c^2+t^2)}\right)^{1/2}x\right) 
\label{efuc}
\end{eqnarray}
Wavefunctions similar to this have been obtained previously\cite{vogel} although not in precisely this form to our knowledge. Here $t_c$ is an arbitrary positive constant with the dimensions of time. $H_n(y)$ are the Hermite polynomials\cite{as}. Square-integrability requires that $n$ be an integer. This wavefunction has been normalised between $\pm \infty$ and the normalisation is constant with respect to time, as it must be. Equation (\ref{efuc}) reduces to a Gaussian wavepacket centred on the origin when $n=0$. This expression may be checked by direct substitution. Henceforth we will retain constants in equations, but all figures will be calculated in atomic units (au) with $\hbar=1$ and $m=1/2$. 

Equation (\ref{efuc}) has one remarkable property. If we look at its form at $t=0$ and define the frequency $\omega=1/t_c$ it becomes
\begin{equation}
\psi(x,t)=\sqrt{\frac{1}{2^nn!}}\left(\frac{m\omega}{\hbar \pi}\right)^{1/4} e^{-m \omega x^2/2\hbar } H_n\left(\left(\frac{m \omega}{\hbar}\right)^{1/2}x\right)
\label{ho}
\end{equation}
Amazingly, this is exactly the form of the eigenfunctions of the harmonic oscillator. Of course this is only true at $t=0$ because our wavefunction evolves according to the free particle Sch\"odinger equation, not the one describing the harmonic oscillator.

This identification helps us because we already know that for the quantum harmonic oscillator the kinetic energy and potential energy operators provide an equal contribution to the total energy of the oscillator. Our Hamiltonian only contains the kinetic energy operator so at $t=0$ and hence at all times
\begin{equation}
E_n=\frac{1}{2}(n+1/2)\hbar \omega=\frac{1}{2}(n+1/2)\frac{\hbar}{t_c}=\frac{1}{2m}<{\hat p}^2>
\end{equation}
where the last equality has been confirmed computationally. 

Of course, it would always be possible to use equation (\ref{ho}) as an initial state of our system and then to integrate the Schr\"odinger equation directly, or to expand it in terms of some basis functions such as plane waves with time dependent coefficients and integrate that. However such a procedure is essentially numerical and would in general require large numbers of basis functions making it rather opaque and unwieldy.
\vskip 5mm
\section{Results}
\subsection{Quantum Mechanical Results and analysis}
\begin{figure}
\includegraphics[width=99mm]{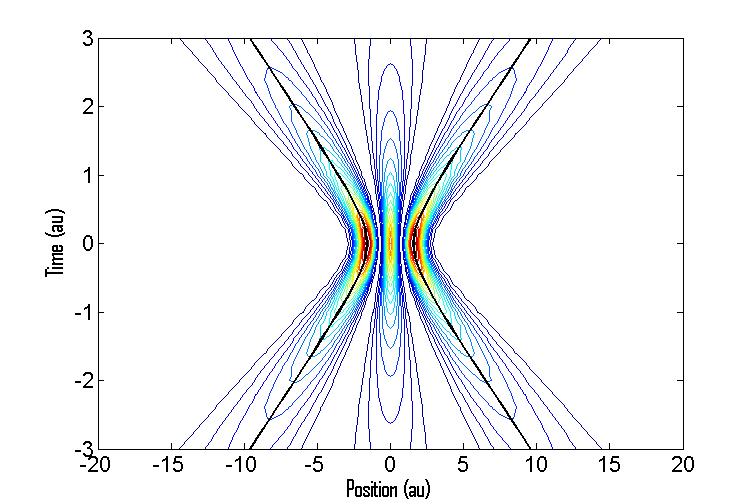}\\
\caption{Contour map of the density associated with the wavefunction of Eqn. (\ref{efuc}) as a function of time and space. This was evaluated for $t_c=1au$ with $n=2$. Superimposed on this are the hyperbolae given by equation (\ref{e9}). }
\label{fig1}
\end{figure}

In Figure 1. we show a contour plot of the probability density associated with this wavefunction as a function of time for $n=2$ and $t_c=1$. The principal effect of $t_c$ is to set the time scale. The wavefunction itself is strongly oscillatory but the oscillatory nature cancels in the probability density to produce three peaks. At all times the probability density is symmetric about $x=0$.  For $t<<0$ the probability density consists of a pair of broad peaks either side of $x=0$ moving towards $x=0$ at approximately constant velocity and a central peak which has its maximum at $x=0$ at all times. As the two outer peaks get close together their contour lines, shown in Figure {\ref{fig1}, describe a curve indicating that the peaks are decelerating and narrowing and at $t=0$ they form two well-localised peaks close to the origin. For $t>0$ this behaviour is reversed with the outer peaks accelerating away from each other and broadening and as $t$ continues to increase the peaks move asymptotically to a constant velocity. The peak at the origin simply broadens as time increases in either the negative or positive directions from zero.

Clearly the path of the outer peaks is hyperbolic and we can find the equation for this path. We can find the maximum of the probability density by differentiating it with respect to $x$. This leads to 
\begin{eqnarray}
2nH_{n-1} \left(\left(\frac{mt_c}{\hbar(t_c^2+t^2)}\right)^{1/2} x\right) = & \cr
\left(\frac{mt_c}{\hbar(t_c^2+t^2)}\right)^{1/2} & x H_n\left(\left(\frac{mt_c}{\hbar(t_c^2+t^2)}\right)^{1/2}x\right)\cr
\end{eqnarray}
as the condition for a wavefunction peak. We can insert the explicit expressions for the Hermite polynomials in here to find the condition for any given $n$. For $n=2$ we find
\begin{equation}
x = \pm \sqrt{\frac{5\hbar}{2mt_c}}(t_c^2+t^2)^{1/2}
\label{e9}
\end{equation}
These two hyperbolae are shown superimposed on the density in Figure 1. Clearly they represent the motion of the wavefunction peaks as a function of time.
\vskip 5mm
\subsection{Semiclassical Analysis}

Accelerating wavefunctions have been observed previously\cite{bb}. In that case the acceleration of the wavefunction was shown to have a classical origin. In this paper we simply perform a similar analysis on the wavefunction (\ref{efuc}) and obtain an analogous result. The key insight found originally and here is that the wavefunction should really be regarded as representing families of particle paths rather than an individual classical particle. The present wavefunction can be regarded as the simplest possible case of this because the initial phase space orbits are simply circular. We will show what we mean by this explicitly as we proceed. 

We will analyse the classical motion using Hamilton's equations. To do this we need an initial position and momentum for the particle. This can be provided by the analogy with the harmonic oscillator at $t=0$. The harmonic oscillator has a Hamiltonian
\begin{equation}
H=\frac{p^2}{2m}+\frac{1}{2}m\omega^2x^2=E
\end{equation}
At $t=0$ we can write the energy of our system in the same way. Writing this down with zero subscripts to indicate that the quantity is valid at $t=0$ only and dividing through by $E$ gives
\begin{equation}
\frac{p_0^2}{\sqrt{2mE_n}}+\frac{m\omega^2}{2E_n}x_0^2=1
\end{equation}
This allows a simple parameterisation of the position and momentum at $t=0$ are
\begin{equation}
p_0(\theta) =\sqrt{2mE}\sin \theta \hskip 10mm x_0 (\theta )=\sqrt{\frac{2E_n}{m\omega^2}}\cos \theta
\label{ini}
\end{equation}
Every different value of $\theta$ in these equations represents the initial conditions for one member of the family of paths. With these initial conditions we can solve Hamilton's equations for a free particle to give 
\begin{figure}
\includegraphics[width=99mm]{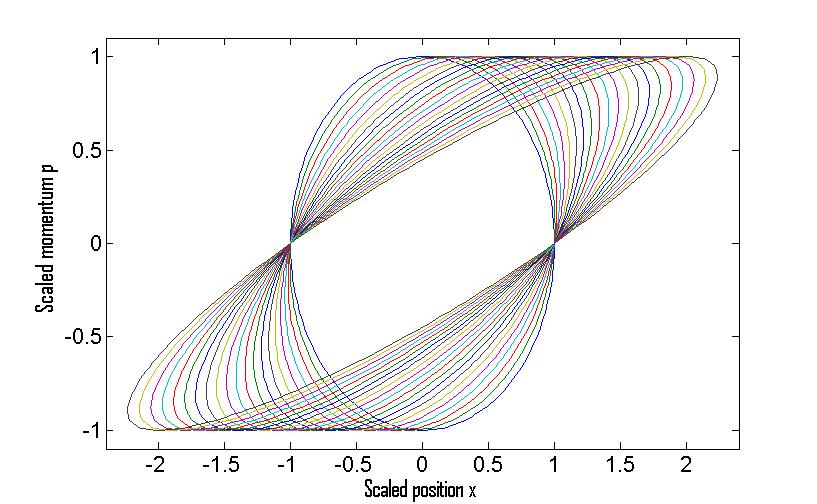}\\
\caption{Phase space curves ($p$ v $x$ for the family of paths described by equations (\ref{ini}) and (\ref{orb}). At $t=0$ this is a circle in phase space which shears into an ellipse as time proceeds.}
\label{fig2}
\end{figure}
\begin{eqnarray}
p & = & p_0(\theta) ={\rm constant} \cr
& \cr
x & = & x_0 (\theta) + pt/m=x_0 (\theta) + p_0(\theta)t/m
\label{orb}
\end{eqnarray}
These equations all represent paths that are straight lines representing uniform motion. As time passes $x$ increases linearly and $p$ remains constant. The phase space curve associated with 
$p_0(\theta)$ and $x_0(\theta)$ is a circle\cite{gold}, but as time passes it shears into an ellipse. This is shown in Figure 2 for times $0\leq t < 2$au. 
\begin{figure}
\includegraphics[width=99mm]{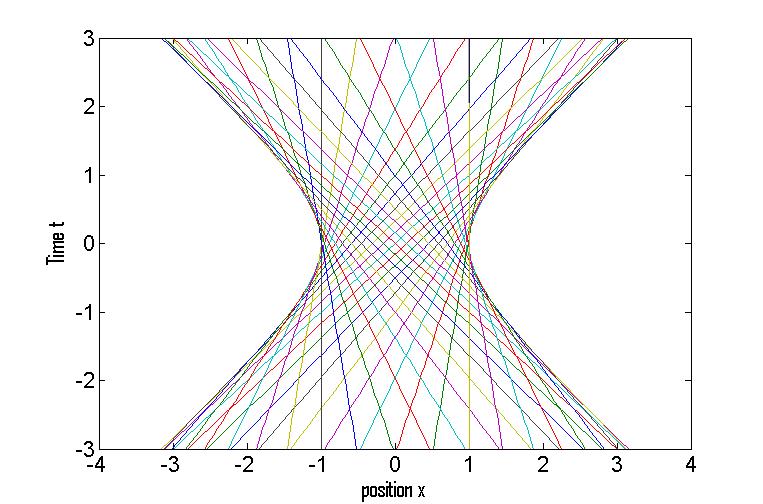}\\
\caption{The family of allowed classical paths for the system described by equation (\ref{orb}). }
\label{fig3}
\end{figure}
Next we plot the possible particle paths described by equation (\ref{orb}) with initial conditions (\ref{ini}). This is shown in figure 3. These paths form a characteristic shape as shown. The edges of this shape form a hyperbolic caustic or envelope function in space-time. An equation for the caustic can be found using standard methods\cite{nye}. i.e. we have to satisfy
\begin{equation} 
x(t) =x(t,\theta ), \hskip 10mm \frac{\partial x(t,\theta )}{\partial \theta}=0
\end{equation}
This is easily done using equations (\ref{ini}) and (\ref{orb}) and results in
\begin{equation}
x(t)=\pm \sqrt{\frac{2E_n}{m}}(t_c^2+t^2)^{1/2}
\end{equation}
For $n=2$ this comes out as 
\begin{equation}
x(t)=\pm \sqrt{\frac{5 \hbar}{2mt_c}}(t_c^2+t^2)^{1/2}
\end{equation}
which is identical to equation (\ref{e9}). Clearly the peak in the space-time representation of the wavefunction in Figure 1. corresponds exactly to the caustic enveloping the classically allowed paths. 
\vskip 5mm
\section{Expectation and Uncertainty}

Lekner has derived the normalisable Airy wavefunction\cite{lek} 
\begin{equation}
\psi(x,t)=Ai[q(x-ut+ivt-\frac{1}{2}at^2)]e^{ima(x-ut-at^2/3)t/\hbar} e^{mv(x-ut+ivt/2-at^2)/\hbar} e^{imu(x-ut/2)/\hbar}
\end{equation}
Here $u$ and $v$ are velocities that are real and $v$ is positive. $a$ is an acceleration. We will consider this wavefunction in its rest frame($u=0$) and compare it with the one derived in this paper. Although checked by us, all results for this wavefunction quoted here were originally published by Lekner and are reproduced here for comparative purposes only. It is instructive to compare the expectation values of position and momentum for the various wavefunctions. This is done in Table I. To make the comparison with the Airy wavefunction as meaningful as possible we have identified $t_c=v/a$. The uncertainties have the usual definitions
\begin{equation}
(\Delta x)^2= <x^2>-<x>^2 \hskip 1cm (\Delta p)^2=<p^2>-<p>^2
\end{equation}
\begin{table}[htdp]
\begin{center}
\begin{tabular}{|c|c|c|c|c|c|}
\hline
$\psi(x,t)$ & $<x>$ & $<x^2>$ & $<p>$ & $<p^2>$ & $(\Delta x \Delta p)^2$ \cr
\hline
Airy & $\frac{v^2}{2a}-\frac{\hbar}{4mv}$ & & 0 &  $\frac{\hbar m}{2t_c}$ &  $\frac{\hbar^2}{4}\left(1+\frac{\hbar}{mv^2t_c}+t^2/t_c^2\right)$\cr
Gaussian ($n=0$) & 0 & $\frac{\hbar}{2m}(t_c+t^2/t_c)$ & 0 & $\frac{\hbar m}{2t_c}$ & $\frac{\hbar^2}{4}(1+t^2/t_c^2)$ \cr
Hermite ($n=1$) & 0 & $\frac{3\hbar}{2m}(t_c+t^2/t_c)$ & 0 & $\frac{3\hbar m}{2t_c}$ & $\frac{9\hbar^2}{4}(1+t^2/t_c^2)$ \cr
Hermite ($n=2$) & 0 & $\frac{5\hbar}{2m}(t_c+t^2/t_c)$ & 0 & $\frac{5\hbar m}{2t_c}$ & $\frac{25\hbar^2}{4}(1+t^2/t_c^2)$ \cr
\hline
\end{tabular}
\end{center}
\label{t1}
\caption{Comparison of the expectation values of the Airy, Gaussian,  and Hermite ($n=1$) and ($n=2$) wavefunctions and evaluation of the uncertainty principle. $<x^2>$ has not been included for the Airy wavefunction because the expression is too long. However it can easily be found from $<x>$ and the uncertainty in equation (\ref{a2})}
\end{table}
For the Airy wavefunction
\begin{equation}
(\Delta x)^2 =\frac{1}{8}\left(\frac{\hbar}{mv}\right)^2 +\frac{\hbar}{2m}(t_c+t^2/t_c)
\label{a2}
\end{equation}
It is easy to show from the table that the uncertainty in position for all the wavefunctions is in accord with the general result\cite{nicola}
\begin{equation}
(\Delta x )^2=(\Delta x)^2_{t=0}+\left(\frac{\Delta p}{m}\right)^2 t^2
\end{equation}
Surprisingly the Airy wavefunction has an identical uncertainty in momentum as the Gaussian wavefunction. The uncertainties in position look very different partially because the Airy wavefunction is not centred on $x=0$. and whether the Airy wavefunction is broader or narrower than the Gaussian wavefunction depends on the size of $v$ when $t_0=v/a$ is kept constant.

\section{Conclusions}

The wavefunction described here illustrates a number of quantum phenomena very clearly. At undergraduate level wavefunctions are normally interpreted in terms of time-independent probability densities. Along with the Airy wave packets derived by Berry and Balazs\cite{bb} and Lekner\cite{lek} the present wavefunction provides a very different perspective where the wavefunction is a description of the family of allowed paths of the particle it describes. The Berry-Balazs wavefunction has a second remarkable property that it does not spread out with time. However it has the drawbacks that it is not square-integrable, is infinite in extent and as a result of this has undefined energy. Along with the Lekner wavefunction the current wavefunction does broaden with increasing (and decreasing) time, but is square integrable and has a well-defined energy, thus making it easier to think about and more appropriate for teaching purposes. This analysis also illustrates the fact that accelerating wavefunctions do not necessarily correspond to accelerating particles as many students might think. Indeed the expectation values of momentum and energy are constant with respect to time. This wavefunction does have the unusual property that it consists of more than one peak that separate as time proceeds, despite the fact that it represents a single particle.However this can be understood if we think about the wavefunction in the semiclassical way discussed here.  Moreover the analysis introduces caustics into quantum theory in an interesting and mathematically simple way. 
\vskip 5mm
\section{Acknowledgements}

I would like to thank Dr L H Ryder for some lively discussions and an anonymous referee whose comments substantially improved this paper.

\bibliographystyle{model1b-num-names}

\begin{thebibliography}{}

\bibitem{sara} S. M. McMurry, {\it Quantum Mechanics}, (Prentice-Hall, 1993)

\bibitem{bb} M. V. Berry and N. L. Balazs, Am J Phys {\bf 47}(3) (1979)

\bibitem{lek} J. Lekner, {\it Eur J Phys} {\bf 30} L43-49, (2009).

\bibitem{lek2} H. Nguyen and J. Lekner,  {\it Applied Mathematics and Computation} {\bf 218}, 10990-10997, (2012).

\bibitem{sivi} G. A. Siviloglou, J. Broky, A. Dogariu, and D. N. Christodoulides, Phys. Rev. Lett. {\bf 99}, 213901, (2007).

\bibitem{hac} S. Hacyan, J. Opt {\bf 13} 105710, (2011).

\bibitem{kam} I. Kaminer, R. Bekenstein, J. Nemirovsky and M. Segev, Phys. Rev. Lett. {\bf 108}, 163901, (2012).

\bibitem{vogel} K. Vogel, F. Gleisberg, N.L. Harshman, P. Kazemi, R. mack, L. Plimak and W. P. Schleich, {\it Chem. Phys.} {\bf 375}, 133-143 (2010) and references therein.

\bibitem{mill} W. Miller Jr. {\it Symmetry and Separation of the Variables}, (Addison Wesley, Boston MA 1977).

\bibitem{kal} E. Kalnins,  {\it SIAM J. Math. Anal}., {\bf 6},340-374 (1975).

\bibitem{as} M. Abramowitz and I. A. Stegun {\it Handbook of Mathematical Functions} (Dover Publications Inc, New York, 1972).

\bibitem{gold} H. Goldstein, C. Poole, and J. Safko, {\it Classical Mechanics}, 3rd Edition, (Addison Wesley, San Francisco, 2002) p.~ 380.

\bibitem{nye} J. F. Nye, {\it Natural Focussing and Fine Structure of Light} (IoP Publishing, Bristol, 1999).

\bibitem{nicola} M. Nicola, {\it Am. J. Phys} {\bf 40} 342, (1972).

\end{thebibliography}

\end{document}